\documentclass[12pt,preprint]{aastex}
\pdfoutput=1
\usepackage{amsmath,longtable,natbib}
\usepackage{lineno}

\shorttitle{MSP gating correlator}
\shortauthors{Roy and Bhattacharyya}

\begin{document}
\def\onoff{{\em on-off}}

\title{Coherently dedispersed gated imaging of millisecond pulsars}

\author{Jayanta Roy\altaffilmark{1}, Bhaswati Bhattacharyya\altaffilmark{2}}
\altaffiltext{1}{National Centre for Radio Astrophysics, Pune 411007, India}
\altaffiltext{2}{Inter-University Centre for Astronomy and Astrophysics, Pune 411007, India}

\begin{abstract} 
Motivated by the need for rapid localisation of newly discovered faint millisecond pulsars (MSPs)
we have developed a coherently dedispersed gating correlator. This gating correlator accounts 
for the orbital motions of MSPs in binaries while folding the visibilities with best-fit 
topocentric rotational model derived from periodicity search in simultaneously generated beamformer output. 
Unique applications of the gating correlator for sensitive interferometric studies of MSPs are illustrated 
using the Giant Metrewave Radio Telescope (GMRT) interferometric array. 
We could unambiguously localise five newly discovered Fermi MSPs in the \onoff\ gated image plane with an 
accuracy of $\pm$1$''$. Immediate knowledge of such precise position allows the use of sensitive 
coherent beams of array telescopes for follow-up timing observations, which substantially reduces the use 
of telescope time ($\sim$ 20$\times$ for the GMRT). 
In addition, precise a-priori astrometric position reduces the effect of large covariances in timing fit 
(with discovery position, pulsar period derivative and unknown binary model), which in-turn accelerates the convergence to
initial timing model. For example, while fitting with precise a-priori position ($\pm$1$''$), timing model converges 
in about 100 days, accounting the effect of covariance between position and pulsar period derivative.
Moreover, such accurate positions allows for rapid identification of pulsar counterpart
at other wave-bands. We also report a new methodology of in-beam phase calibration using the \onoff\ gated image of 
the target pulsar, which provides the optimal sensitivity of the coherent array removing the possible 
temporal and spacial decoherences. 

\end{abstract}

\vskip 0.6 cm

\section{Introduction}
\label{sec:intro}

The ongoing sensitive surveys (e.g. Fermi directed radio searches, \cite{Ray12}; 
Green Bank drift scan Survey\footnote{http://www.as.wvu.edu/$~$pulsar/GBTdrift350/};
Green Bank North Celestial Cap (GBNCC) survey\footnote{http://arcc.phys.utb.edu/gbncc/}, 
High Time Resolution Universe (HTRU) survey at Parkes, \cite{Keith10}; Pulsar Arecibo 
L-Band Feed Array (PALFA) survey, \cite{Cordes06}) have discovered a number of intriguing
fainter millisecond pulsars (MSPs). As a result in last three years the population of Galactic disk MSPs is increased by 
about 45\%\footnote{http://astro.phys.wvu.edu/GalacticMSPs/GalacticMSPs.txt}. 
Sensitive follow-up studies of these newly discovered MSPs using coherent beams of array 
telescopes, or at higher frequencies using single dishes, are hindered by the large 
uncertainties associated with the discovery positions. For example, discovery uncertainties
range from 40$'$ for the GMRT at 322 MHz (GMRT$-$322) to 4$'$ for the Arecibo at L-band.
Sensitive coherent array follow-up observations significantly reduces the use of array 
telescope time ($\sim$ 20 $\times$ for the GMRT), which is important as arrays are the 
future for large radio telescopes. Such coherent array observations improves the uncertainties 
in time-of-arrivals (TOAs) and allows to generate much closely spaced TOAs in order to avoid 
ambiguous phase connection while timing the binaries with shorter orbits.
Traditionally long-term pulsar timing programs are used to reduce such positional uncertainties, 
requiring significant amount of observing time. Simultaneous timing fit with discovery position 
and unknown binary parameters can be affected from large covariances, specially for long period 
binaries. In addition, the covariance between position and pulsar period derivative ($\dot{P}$) limits 
the convergence of timing fit even with known binary model. The effects of such large covariances in 
timing fit can be minimised with precise a-priori astrometric position. 

Being compact objects pulsars are effectively seen as point sources in interferometric imaging. 
Pulsars specially MSPs are weak radio sources in continuum image plane, having fluxes in the range of few mJy even 
at lower frequencies. Identification of pulsar counterpart in the continuum image can also be confused 
with the other sources in the field of view. Considering the narrow duty cycle, 3\%$-$10\% \citep{henry69}, the detection significance of a 
pulsating point source can be largely improved by removing the off-pulse noise. This is achieved through pulsar 
gating, where the continuum image is sampled synchronously with the pulsed signal to generate a number of 
gated images for different pulse phases (e.g. pulsar gating at the 
ATCA, \cite{Lazendic99}).
Background sky subtracted \onoff\ gated image allows to unambiguously identify the location of pulsed 
emission \citep{Camilo00}. However, such precise position determination using gating is hitherto not been reported 
for MSPs. High time resolution gating requirements for MSPs, with the gating window $\sim$ 100 $\mu$s, could be computationally
challenging. More importantly for MSPs, since period and intrinsic pulse width are quite small, dispersion correction 
is required to be done (unlike normal pulsars) before such high time resolution gating, in order to 
account for large dispersion delay. 
For example, considering a MSP with a dispersion measure (DM) of 30 pc/cc, the dispersion delay at 322 MHz GMRT frequency 
across 32 MHz band is $\sim$ 247 ms, which is many times of MSP period. In earlier studies gating has been done 
after correlation using incoherent dedispersion. But incoherent dedispersion smearing across frequency channels can be 5 times 
larger than intrinsic pulse width (considering a 2 ms MSP at 30 pc/cc DM with 5\% duty cycle for GMRT$-$322), significantly 
reducing the number of effective gates due to pulse smearing. Thus reconstructing intrinsic pulse width with coherent 
dedispersion \citep{Hankins75} will be beneficial while performing gating with larger bin numbers requiring sufficient time resolution. 
In addition, since majority of MSPs are in binary system, the effect of orbital motion on pulsar period has to be accounted for in 
MSP gating correlator. 

Alternatively localisation of newly discovered pulsars can be achieved by forming multiple coherent beams covering the 
primary beam of the telescope. For example the grating response of a linear array like the WSRT can be used for localisation
with an accuracy of few arcminutes \citep{Rubio-Herrera12}. Whereas with continuum imaging followed by the multi-pixel 
beamformer using nonlinear array like the GMRT \citep{Roy12} we have achieved the positional accuracy of few arcseconds 
(determined by the synthesized beam of the array). But this method is not efficient to localise relatively 
fainter MSPs having very low detection significance in continuum image plane.

In order to improve the detection significance in image plane and to achieve positional accuracy of the order of 1$''$, 
we have developed a coherently dedispersed MSP gating correlator at the GMRT. 
In addition to aperture synthesis at moderate time constants ($\sim$ seconds) and high 
time resolution incoherent and coherent beam formation ($\sim$ 30 $\mu$s), the GMRT Software Back-end (GSB; \cite{Roy10})
is equipped to stream the raw base-band samples from all the antennas to array of storage disks. In this paper we 
describe design and implementation of the MSP gating correlator using raw Nyquist sampled base-band data and its application
to obtain the precise locations of five newly discovered MSPs from the Fermi directed searches. In addition we demonstrate 
an unique application of gated imaging, in using pulsar as a phase calibrator to achieve the optimal sensitivity for the 
coherent array.

\section{MSP gating correlator} 
\label{sec:gating_corr}

Design of the MSP gating correlator that can increase the signal-to-noise (S/N) of a pulsar in image plane by a 
factor of 3 to 6 (approximately proportional to the inverse square root of duty cycle) is described below. 

(a) Coherent dedispersion :
Since baseline based incoherent dedispersion performed before folding the visibility time-series at the best-fit
topocentric model is not adequate for the study of MSPs at lower frequencies, we implement antenna based coherent 
dedispersion. Antenna based coherent dedispersion ($\sim$ N operations) can also be computationally favorable compared to baseline based 
incoherent dedispersion ($\sim$ N$^2$ operations) for the future arrays with large number of elements (N). 
Coherent dedispersion module is running on the recorded raw base-band data prior to correlation, and is parallelized
over telescopes. The dedispersed voltage samples are written to a shared memory ring buffer.
 
(b) Correlation and folding :
Dispersion corrected raw voltage samples read concurrently from the shared memory are correlated to generate the high time 
resolution visibility time-series. These visibilities are fed into a gating module which bins the data in multiple gates. 
For a given MSP, while folding using best-fit topocentric rotational model, number of gates and intermediate time 
resolution of dedispersed visibilities, are empirically adjusted to obtain optimal S/N.
In order to retain this optimal S/N in a 1 hr gated image for a 2 ms MSP (considering gate width approximately equal to pulse width), 
fractional period error per rotation ($\dot{P}$) is required to be below $\sim$ 10$^{-13}$, i.e. MSP period (in ms) is required 
to be correct up to 7 significant digits. 
Folding with a fixed topocentric period is sufficient for imaging isolated or loosely coupled binary MSPs. 
However, we have parameterised the rotational model to include period ($P$), acceleration ($\dot{P}$) and 
jerk ($\ddot{P}$) for folding MSPs in tighter orbits. For newly discovered MSPs this topocentric rotational model is derived from 
PRESTO \citep{Ransom02} based periodicity search for the same observation using simultaneously generated incoherent beamformer 
output. Whereas for MSPs with known ephemeris we have used TEMPO\footnote{http://www.atnf.csiro.au/research/pulsar/tempo} based 
predictions. The folded visibilities on each gate are then integrated up to a time resolution of 16 seconds and the complex correlation 
output from each gate are finally written to disks for further processing.
 
(c) \onoff\ gating :
The on-pulse gate is also identified using the coherent, or incoherent where the position of the MSP is poorly known, beamformer 
output, as the arrival time of the pulse is unknown a-priori for new pulsars.
The gates containing the pulsed emission are grouped to form the on-pulse data, and the off-pulse data is formed with same 
number of gates covering the off-pulse phases. In order to unambiguously identify location of pulsed emission we have 
generated \onoff\ visibility bin. In addition, by subtracting the nearby off-gate from the on-gate underlying systematics 
generated from instrumental effects as well as from radio frequency interferences are canceled out, resulting in further 
improvement in noise statistics.

(d) Calibration and imaging :
Coherently dedispersed folded \onoff\ visibility data are then flagged for removing the outliers and 
calibrated for solving the complex gains \citep{Prasad11}. Calibrated visibilities are imaged and deconvolved using standard 
imaging package AIPS \footnote{http://www.aips.nrao.edu}. MSP is the {\it only} source in this \onoff\ gated image,  considering that
on and off gates contain equal flux for other sources.  
 
\section{Localisation of MSPs using gating correlator} 
\label{sec:application}
MSP gating correlator is efficient in localising fainter MSPs, where fluxes are around 1 mJy resulting in 
very low detection significance on continuum image plane (rms noise in 1 hr at GMRT$-$322 $\sim$ 500 $\mu$Jy). 
Positional accuracy achieved from \onoff\ gated image depends upon hour angle of the observations and S/N of pulsar detection.
Considering GMRT$-$322 synthesized beam (FWHM) $\sim$ 10$''$, the positional accuracy of a MSP scales according to FWHM/(2$\times$S/N).
For a typical S/N of 5 (Table. \ref{gating_param}), an accuracy $\pm$1$''$, is determined by the AIPS task JMFIT. 
Such a-priori astrometric accuracy accelerates the convergence in pulsar timing. The newly discovered MSPs have large
uncertainties in the a-priori position, requiring timing span of the order of a year to overcome the
effect of covariance between position and $\dot{P}$. However, while fitting with more precise
a-priori position, having $\pm$1$''$ uncertainty, the positional accuracy of $\sim$ 1 mas and
a convergence in detection of $\dot{P}$ (with $\Delta\dot{P}$/$\dot{P}$ $\sim$ 0.04) can be achieved 
in only about 100 days.

We performed gated imaging for five MSPs discovered in Fermi directed radio searches. Among those, PSR J1120$-$3618, J1646$-$2142 
and J1828$+$0625 were discovered by us in GMRT$-$322, whereas PSR J1207$-$5050 in GMRT$-$607 \citep{Ray12} and PSR J1551$-$0658
in GBT$-$350 \citep{Bangale12}. The parameters of these MSPs such as period, DM and mean flux are listed in Table \ref{psr_param}. 
Quoted mean flux is obtained using the simultaneous incoherent beamformer output. For PSR J1207$-$5050 the gating 
observations were done at 607 MHz while rest of the MSPs were observed at 322 MHz.

\begin{figure}[h!]
\epsscale{0.70} \plotone{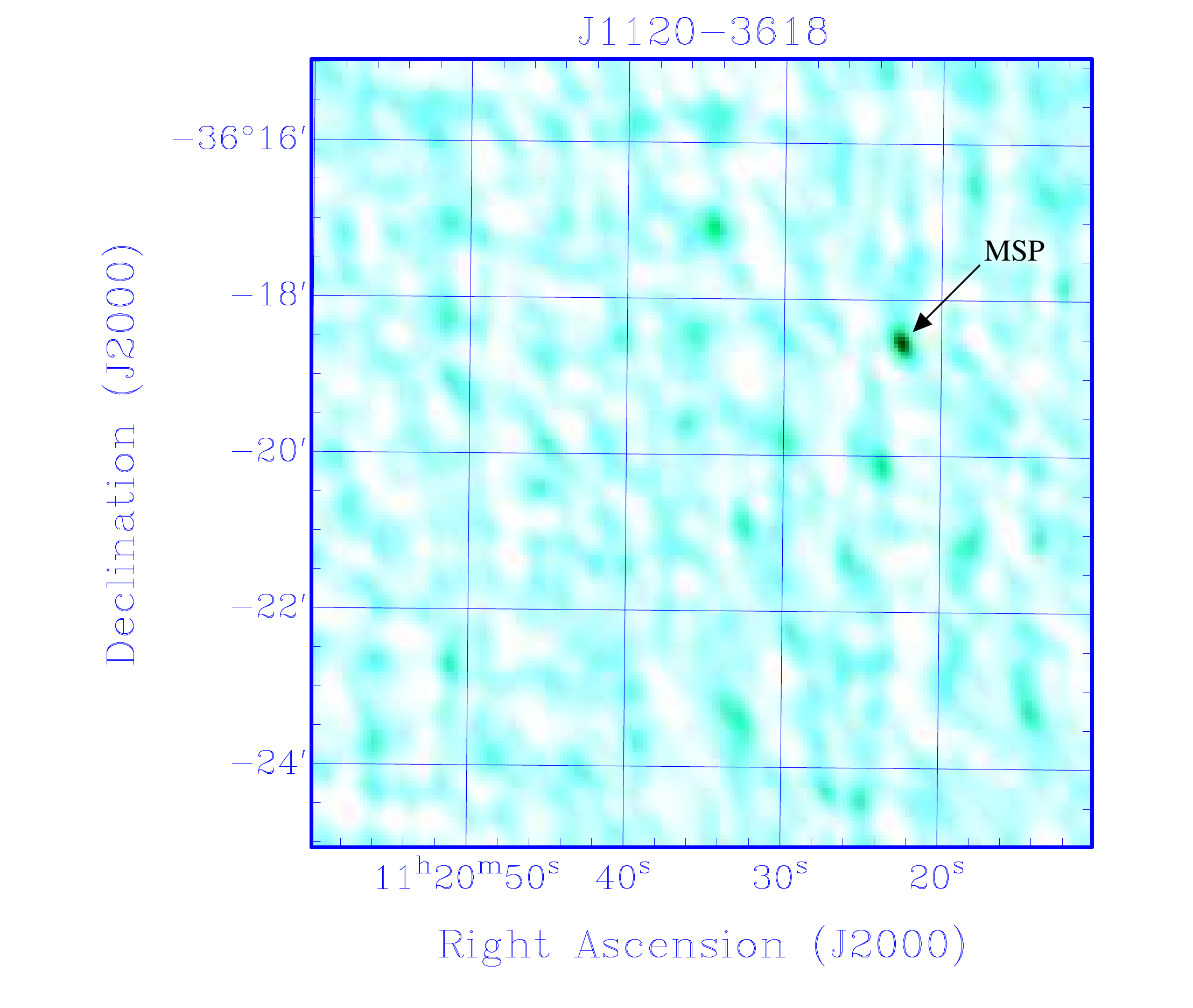}
 \caption{Localization of the PSR J1120$-$3618 using MSP gating correlator. The MSP (marked in the image) is detected in 
the \onoff\ gated image at 57\arcmin~offset from pointing centre\label{fig:1120-3618}.}
\end {figure}

PSR J1120$-$3618 is a serendipitously discovered MSP in the GMRT$-$322 field-of-view of known Fermi MSP J1124$-$3653. This is a very faint MSP
(mean flux $\sim$ 300 $\mu$Jy) in a relatively tighter orbit. PRESTO based search pipeline reports an acceleration 
equal to 0.1 m/s$^2$, corresponding to a significant measurement of $\dot{P}$ equal to 2$\times$10$^{-12}$ s/s. This introduces 
a period (in ms) error at 1 part in 10$^5$ over 3 hrs of observation, whereas for optimal folding an accuracy up to 1 part in 10$^6$
over 3 hrs is needed. Thus accounting for acceleration is required to retain the S/N of this MSP in \onoff\ gated image plane. 
The pulse phase is binned in 11 gates (gating window $\sim$ pulse width) at the intermediate time resolution (491.52 $\mu$s) of the 
dedispersed visibilities. The \onoff\ gated image for this MSP is shown in Fig. \ref{fig:1120-3618}. 
Interestingly this MSP is located at 57\arcmin~offset from the pointing center, which is out-side the GMRT$-$322 beam-width 
($\sim \pm$ 40\arcmin).
To localise the MSP we have performed multi-faceted \citep{Perley99} gated imaging which is also corrected for primary beam effect as 
the pulsar is located at the edge of the beam. In order to obtain the gated image of the full field-of-view, separate facet images 
are interpolated and averaged onto a larger grid using the AIPS task {\it FLATN}. A 10\arcmin$\times$10\arcmin~facet of 
the \onoff\ gated image shows the MSP (Fig. \ref{fig:1120-3618}) with 5$\sigma$ detection significance, resulting in 
gated flux of 1.2 mJy which is 4$\times$ the mean flux as expected from gating. The parameters related to gated imaging 
are listed in Table. \ref{gating_param}. Sensitive coherent beam is formed towards the pulsar location by steering the 
phase center and an expected sensitivity improvement $\sim$ 4$\times$ of incoherent array detection is achieved \citep{Roy12}. 

\begin{figure}[h!]
\epsscale{1.00} \plotone{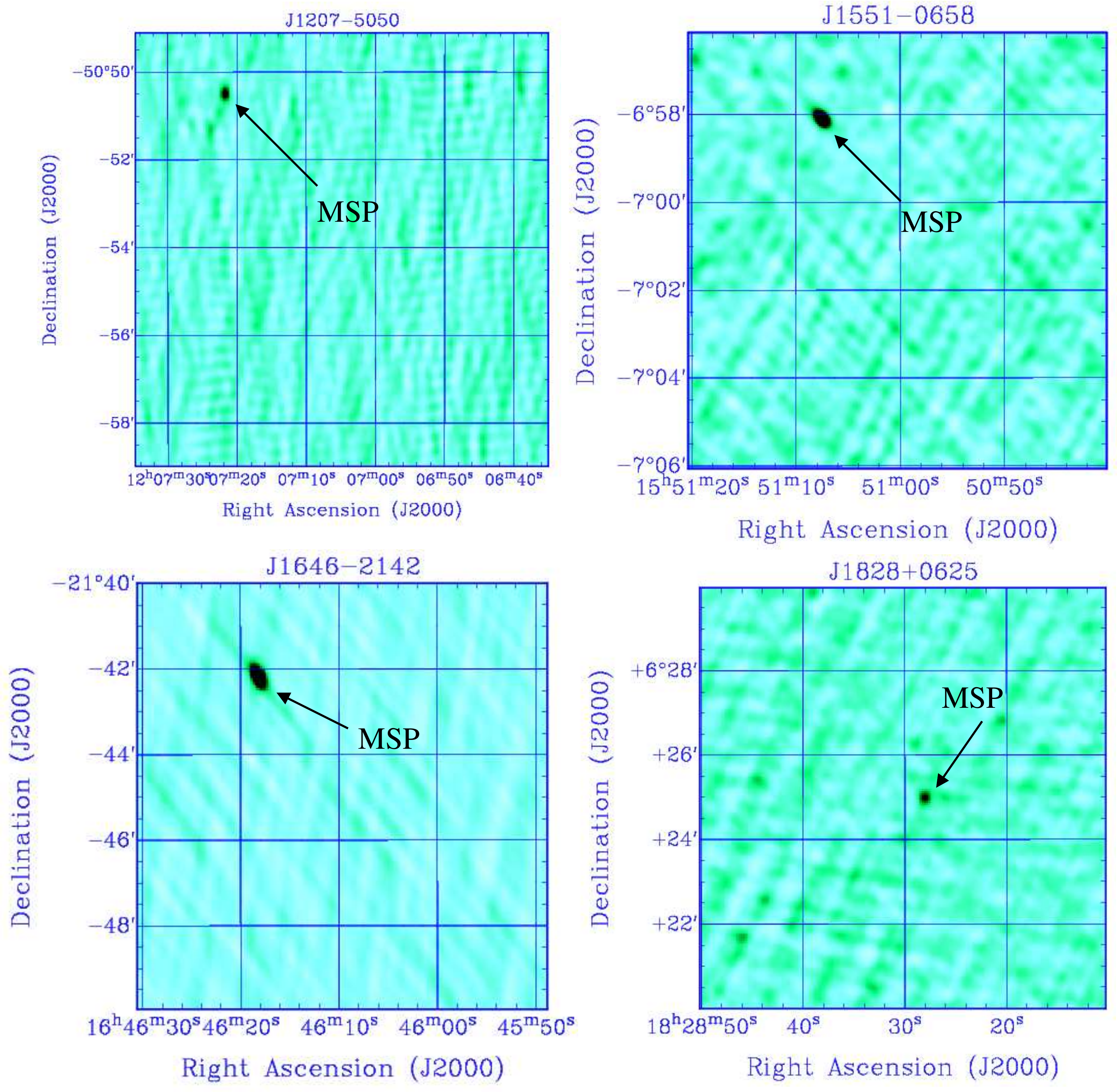}
 \caption{The \onoff\ gated images for PSR J1207$-$5050, J1551$-$0658, J1646$-$2142 and J1828$+$0625. All the MSPs are marked in the 
respective 10\arcmin$\times$10\arcmin~facet images\label{fig:4_MSP}.}
\end{figure}

PSR J1207$-$5050, J1551$-$0658, J1646$-$2142 and J1828$+$0625 are also successfully localized in their respective \onoff\ 
gated images (shown in Fig. \ref{fig:4_MSP}). Table. \ref{gating_param} lists the related parameters.
PSR J1207$-$5050 and J1646$-$2142 are found within the error radius of the associated Fermi sources. Whereas PSR J1551$-$0658 
and J1828$+$0625 are located at an offset of 20\arcmin~and 26\arcmin~respectively from their pointing centers (outside the error radius), 
indicating that these MSPs are not likely to be associated with the corresponding gamma-ray sources. Large primary beam of the GBT$-$350 
and large incoherent beam of the GMRT$-$322 have allowed such serendipitous discoveries. The gated fluxes for all these 
four MSPs are 3 to 6 times of the respective mean fluxes as expected from the S/N improvement by gating.

\begin{table}
\begin{center}
\caption{Parameters of the concerned MSPs}
\vspace{0.3cm}
\label{psr_param}
\begin{tabular}{|l|c|c|c|c|c|c|c|c|c|c|c|c|c|c|c|c}
\hline
PSR     	&Period &Dispersion    	&Mean   \\
        	&(ms)	&measure (pc/cc)&flux$^a$ (mJy)   \\\hline
J1120$-$3618	&5.55	&45.1		&0.3   	       \\
J1207$-$5050	&4.84   &50.7           &0.2            \\
J1551$-$0658	&7.09   &21.6           &1.0            \\
J1646$-$2142	&5.85   &29.7           &2.1            \\
J1828$+$0625	&3.63   &22.4           &1.3            \\\hline
\end{tabular}
\vspace{0.3cm}\\
$a$ : Mean flux of PSR J1207$-$5050 is measured at 607 MHz, rest are in 322 MHz \\
\end{center}
\end{table}
\begin{table}
\begin{center}
\caption{Parameters related to gated imaging}
\vspace{0.3cm}
\label{gating_param}
\begin{tabular}{|l|c|c|c|c|c|c|c|c|c|c|c|c|c|c|c|c}
\hline
PSR     &Gated J2000      &Offset from     &Number     &Observing      &Gated      &Gated \\
        &position 	  &pointing        &of gates   &duration       &flux$^a$   &SNR   \\ 
  	&(Errors in \arcsec)&centre	   &	       & (min)          &(mJy)      &     \\\hline
J1120$-$3618&11$^\mathrm{h}$20$^\mathrm{m}$22\fs405 (1\farcs1);&57\arcmin	 & 11	   &180  &1.2        &5   \\
            &-36\degr18\arcmin32\farcs17 (2\farcs2)      	&                &         &   &           &    \\
J1207$-$5050&12$^\mathrm{h}$07$^\mathrm{m}$21\fs811 (0\farcs4); &6.2\arcmin      & 10      &120  &1.1        &6    \\
            &-50\degr50\arcmin30\farcs27 (1\farcs4)           &            	 &         &   &           &    \\
J1551$-$0658&15$^\mathrm{h}$51$^\mathrm{m}$07\fs215 (0\farcs6);&20\arcmin      & 14      &60  &5.8        &11    \\
            &-06\degr58\arcmin06\farcs51 (0\farcs6)	         &               &         &   &           &  \\
J1646$-$2142&16$^\mathrm{h}$46$^\mathrm{m}$18\fs127 (0\farcs9); &10\arcmin      & 12      &60  &11.3       &11  \\
            &-21\degr42\arcmin08\farcs96 (1\farcs4)            &            	 &         &  &           &    \\
J1828$+$0625&18$^\mathrm{h}$28$^\mathrm{m}$28\fs030 (1\farcs0);  &26\arcmin      & 15      &45  &3.5        &6    \\
            &06\degr25\arcmin00\farcs52 (1\farcs3)		 &               &         &  &           &   \\\hline
\end{tabular}
\vspace{0.3cm}\\
$a$ : Gated flux of PSR J1207$-$5050 is measured at 607 MHz, rest are in 322 MHz \\
\end{center}
 \end{table}


\section{In-beam phase calibration using pulsar} 
\label{sec:inbeam_cal}
The sensitive coherent beamformer allows study of high time resolution temporal variations of pulsars. 
In order to form coherent beam using an array telescope like the GMRT, antenna based complex 
gains (amplitudes and phases) need to be solved using recorded visibilities on a calibrator source. 
The optimal baseline length over which the array can be coherently added is limited by the perturbations in 
ionospheric phases, which are more severe at lower frequencies. Coherent array sensitivity degrades 
with time due to the temporal decoherences caused by instrumental as well as ionospheric phase 
fluctuations. Such degradation can be reduced with interleaved calibrator observations.  
However, applying the phases derived from a distant calibrator into a pulsar field can cause some more 
decoherences due to underlying different ionospheric inhomogeneities \citep{Thompson86}. 
We have used the \onoff\ gated image of the target pulsar as a sky model to solve for antenna based residual 
stochastic phase errors (affecting the data on short time scale) as well as the broad band phase offsets applying 
self-calibration in AIPS. This process 
produces a set of phase solutions with time, written in SN table generated by AIPS. While forming the coherent array, 
these residual phase solutions are recursively applied, in addition to initial narrow-band phase corrections 
derived from the calibrator visibilities. Background sky subtracted \onoff\ gated image 
of the target pulsar provides a better model to solve for phases since the effects of instrumental as well as RFI artifacts 
are canceled out. A pulsar with about 10 mJy flux density allows this in-beam calibration process to converge (ensuring more 
than 3$\sigma$ detection in each frequency channel during calibration with 10 minutes cadence). This in-beam calibration is 
the optimal way of coherently adding all the working antennas of an array telescope. 

\begin{figure}[h!]
\epsscale{1.00} \plotone{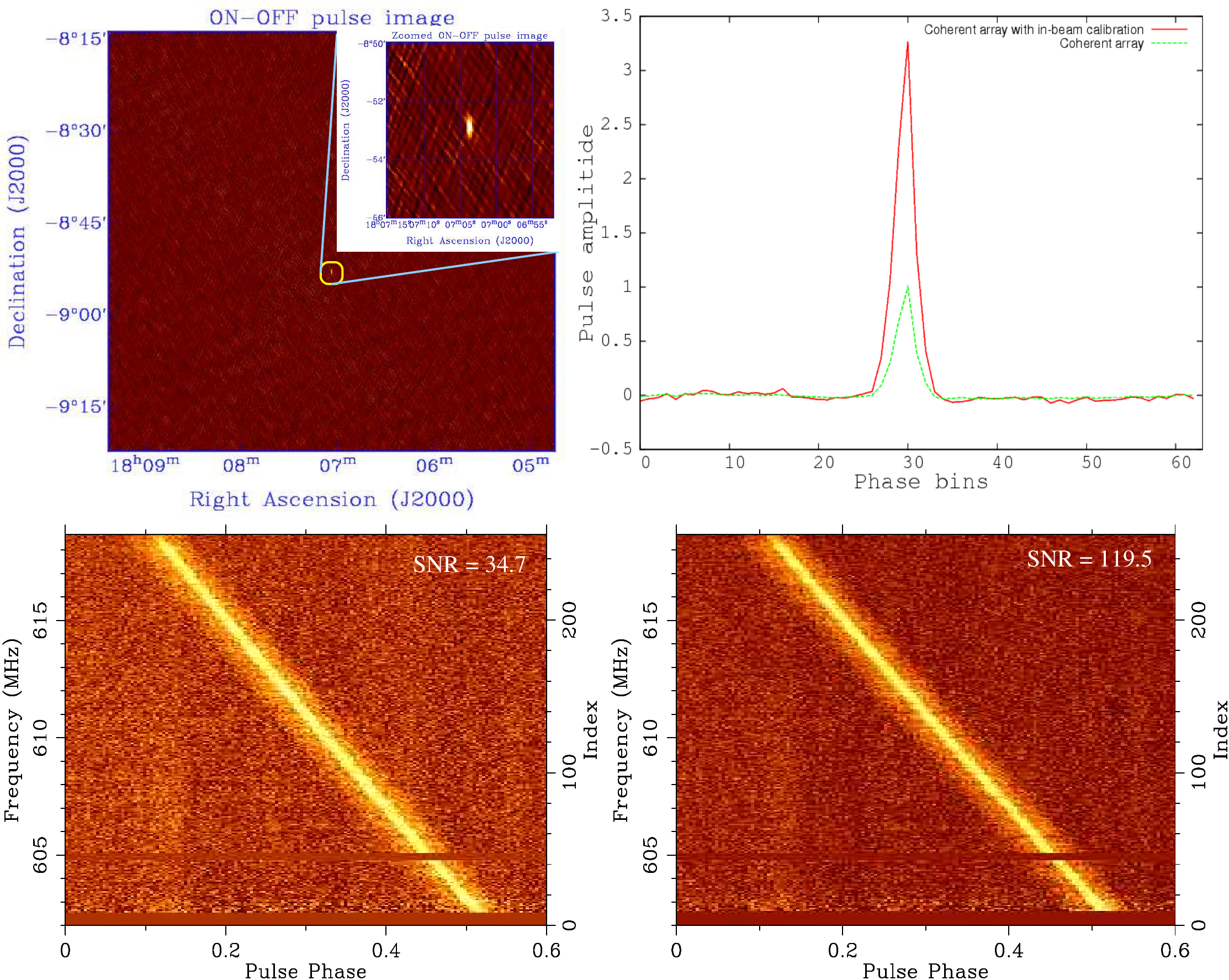}
 \caption{In-beam phase calibration using PSR B1804$-$08. Upper left panel shows the \onoff\ gated image with a zoom around the pulsar. 
Upper right panel shows the coherent array sensitivity improvement in the folded pulse profile using in-beam phase calibration. The 
dispersed pulse phase with frequency is shown in the bottom panel. The similar sensitivity improvement with reduction of spectral
noise (S/N are mentioned on the right corners of the respective plots) is seen on the right panel with respect to the left panel\label{fig:phase_cal}. }
\end{figure}

We have generated \onoff\ gated image for PSR B1804$-$08 using 1 minute of base-band data, which is shown in upper left 
panel of Fig. \ref{fig:phase_cal}. With the in-beam calibration a sensitivity improvement of 3.5$\times$ compared 
to the conventional coherent array is seen in the folded profile (upper right panel of Fig. \ref{fig:phase_cal}). 
This improvement includes contributions from the increased coherence length of the array (achieved from 50\% increase in 
number of antennas) and better modeling of phase errors.
Since the spectral voltages from the antennas are optimally added in phase, the sensitivity improvement can also be 
visualized as a reduction in the spectral noise in the dispersed pulse phase versus frequency plot (bottom panel of 
Fig. \ref{fig:phase_cal}) generated with DSPSR \citep{Straten10}. 
In addition, the temporal decoherence affecting the long observing scans of target pulsars can be avoided by recursive 
in-beam calibrations at short time scale, without slewing to a distant calibrator location. Use of pulsar as an in-beam 
phase calibrator can introduce a lateral shift in the image domain while pulsar is not at the phase centre. This can be 
calibrated using known position of any in-field catalogue source allowing to perform an astrometric measurement of pulsar. 

\section{Summary and future scope} 
\label{sec:summary}
We report design and implementation of a coherently dedispersed MSP gating correlator which 
accounts for a full timing model including orbital motion while folding the visibility time-series.
In this paper we have unambiguously determined the precise positions, with $\pm$1$''$ accuracy, for 
five newly discovered MSPs using the \onoff\ gated imaging. 
Locatisations of such relatively fainter MSPs are greatly benefited from the significant enhancement in S/N on the gated image plane compared to the 
normal synthesis observations. 
Knowledge of such accurate positions in immediate observations after discovery allows follow-up observations 
with sensitive coherent array, substantially reducing the use of telescope time by an order of magnitude. 
Inaccuracy of the astrometric model, associated with large uncertainties in discovery positions, increases the length of data 
span required to reduce the effect of covariance between position and $\dot{P}$ in timing fit. While determining an unknown 
binary model, large covariance between position and the binary parameters, specially for long 
period binaries, can influence the timing fit, which can be minimised with accurate astrometric position ($\sim\pm$1$''$).
In addition, for pulsars located near the ecliptic plane, the astrometric inaccuracy in ecliptic latitude 
from the timing fit is much larger \citep{Lorimer04} indicating the need for interferometric gated observations.
This accurate localisation will also facilitate the search for pulsar counterpart and possible binary companions at optical and X-ray.
Current astrometric positional accuracy ($\pm$1$''$) is decided by array size and S/N of detection.
Application of MSP gating correlator in the VLBI \citep{deller09} 
and the SKA 
scale improves this a-priori astrometric accuracy (e.g. at 1.4 GHz, 3000 km SKA baseline can give astrometric 
precision as 15 $\mu$as \citep{Smits11}).

In addition we have used the \onoff\ gated image of the target pulsar to correct the residual 
phase errors in order to avoid degradation of coherent array sensitivity caused by decoherences from instrumental and 
ionospheric phase fluctuations. Such in-beam phase calibration using a target pulsar ensures optimal sensitivity of the coherent 
array. We believe that this methodology will be very fruitful for recently commissioned and up-coming array telescopes 
(e.g. LOFAR, \cite{de Vos09}; MWA, \cite{Lonsdale09}; ASKAP, \cite{Johnston08}; MeerKAT, \cite{Jonas09}).
  
Even though the design is primarily motivated by the requirement of localisation of newly discovered Fermi MSPs, this gated 
imaging has the potential to unfold some other interesting properties of MSPs. Firstly, the MSP gating correlator will allow 
to perform independent measurement of parallaxes and proper motions even for relatively fainter binary MSPs, which will 
largely benefit pulsar timing and will probe the interstellar medium (ISM) at various line-of-sight (\cite{McGary01}, \cite{Lorimer04}, 
\cite{Smits11}). Secondly, the study of un-pulsed emission associated with 
pulsar wind nebulae (PWN), can be performed by imaging the pulsar field when pulsed emission is off \citep{Gaensler00}. Angular 
extent of PWN for high energy pulsars in low density ISM can be as large as few arcminutes (e.g. \cite{Frail94}), which can be 
detected with low frequency gated imaging using this MSP gating correlator. 
Finally, pulsar itself may have a weak off-pulse emission, having magnetospheric origin, coming from close to light cylinder 
(\cite{Perry85}, \cite{Basu12}). Origin of possible off-pulse emission from MSPs can be probed with the gated images as 
a function of pulse longitude and any possible co-location with the gamma-ray emission region will be interesting to investigate.
Thus the design of coherently dedispersed MSP gated imaging assures broader scientific returns, while also 
having importance for SKA applications. 
 
\section{Acknowledgments}

We thank computer group at GMRT and NCRA. We thank Dr. David Thompson and Dr. Paul Ray 
for helping us to join PSC and Prof. Yashwant Gupta and Prof. Dipankar Bhattacharya 
for being members of the GMRT team. We also thank Prof. Ue-Li Pen for insightful 
discussions and Dr. Maura Mclaughlin for inspiring to work on J1551$-$0658.  We thank
the anonymous referee for comments that improved the quality of the paper.
We acknowledge support of telescope operators. The GMRT is run by the National Centre for Radio Astrophysics of the 
Tata Institute of Fundamental Research.


\end{document}